\documentstyle[12pt]{article}
\newcommand{\cl}{\centerline}

\newcommand\beq{\begin{equation}}
\newcommand\eeq{\end{equation}}
\newcommand\bea{\begin{eqnarray}}
\newcommand\eea{\end{eqnarray}}
\begin{document}

\begin{titlepage}
\setlength{\textwidth}{5.0in}
\setlength{\textheight}{7.5in}
\setlength{\parskip}{0.0in}
\setlength{\baselineskip}{18.2pt}
\hfill YITP-SB-01-12, CITUSC/01-013
\vskip 0.3cm
\cl{\Large{{\bf Partition functions and Jacobi fields}}}\par  
\cl{\Large{{\bf in the Morse theory}}}\par
\vskip 0.8cm
\begin{center}
{Soon-Tae Hong$^{*}$}
\end{center}
\begin{center}
{\it C.N. Yang Institute for Theoretical Physics,}\par
{\it State University of New York at Stony Brook, Stony 
Brook, NY 11794, USA}\par
{\it and}\par
{\it CIT-USC Center for Theoretical Physics,}\par
{\it University of Southern California, Los Angeles, CA 90089-2535, USA}
\par
\end{center}
\cl{\today}
\vskip 0.5cm
\vfill
\cl{\large Abstract}
\begin{quotation}
\noindent
We study the semiclassical partition function in the frame work of the Morse
theory, to clarify the phase factor of the partition function and to relate it 
to the eta invariant of Atiyah.  Converting physical system with potential 
into a curved manifold, we exploit the Jacobi fields and their corresponding 
eigenvalue equations to be associated with geodesics on the curved manifold 
and the Hamilton-Jacobi theory.   
\vskip 0.2cm
\noindent
{\footnotesize $^{*}$ On leave from Department of Physics, Sogang 
University, Seoul, Korea}
\noindent
{\footnotesize $^{}$E-mail address: sthong@ccs.sogang.ac.kr}

\end{quotation}
\par
\end{titlepage}

\newpage

Nowadays there have been considerable discussions concerning the 
topological invariants such as the Euler characteristics, the Hopf invariant 
and so on in mathematical physics.  Moreover, the eta invariant of Atiyah 
has been for instance studied by Witten in quantum field theory 
associated with the Jones polynomial and knot theory~\cite{witten89} and even 
in hadron physics such as the chiral bag model~\cite{hong93}.  Recently, 
the Yamabe invariant~\cite{anderson97,gibbons99} are also investigated in 
general relativity associated with the topology of boundary surface of black 
holes with nontrivial higher genus.

On the other hand, since Feynman proposed the path integral formalism in 
1948~\cite{feyn48}, there have been tremendous developments in quantum field 
theory.  Especially, the partition functions in the path integral scheme have 
become crucial in investigating many aspects of recent theoretical physics.  
In fact, Morette studied the partition functions 
semiclasically long ago, but her expression for its phase factor possesses 
somehow subtleties~\cite{morette51}.  

In this paper we reformulate the semiclassical partition function in the 
frame work of the Morse theory of differential geometry, to clarify the phase 
factor of the partition function and to relate it to the eta invariant of 
Atiyah~\cite{atiyah75,witten89}.  To do this, we will convert the physical 
system with potential into a curved manifold, on which we will use the 
Jacobi fields and their corresponding eigenvalue equations associated with 
the geodesics on the curved manifold, and the Hamilton-Jacobi theory.   

Now we consider a particle in a conserved physical system with constant energy 
$E$, which consists of the kinetic and potential energies given as 
$T=m\delta_{ij}v^{i}v^{j}/2$ with $v^{i}=dx^{i}/d\tau$ and $V=V(x^{i})$, 
respectively.  Converting the potential energy into curvature of the 
target manifold, one can obtain action of the particle of the form
\footnote{On the curved manifold $M$, we will use the abstract index 
notation $a, b$ which can be converted into the component notation whenever 
we wish.  Here one notes that, without loss of generality, $V(x^{i})$ can be 
chosen to vanish at starting point at $\tau_{1}$.  See Ref.~\cite{wald84} for 
the notations and conventions used here as in Eq. (\ref{riemann}).} 
\beq
S=\int_{\tau_{1}}^{\tau_{2}}{\rm d}\tau~(g_{ab}v^{a}v^{b})^{1/2},
\label{action}
\eeq
to yield a curved three-dimensional manifold $M$ with metric 
\beq
g_{ab}=\frac{m(E-2V)^{2}}{2(E-V)}\delta_{ab},
\label{metric}
\eeq
which does not have any singularities since the denominator of Eq. 
(\ref{metric}) is positive definite.  With the metric $g_{ab}$ in mind, one 
can define a unique covariant derivative $\nabla_{a}$ satisfying 
$\nabla_{a}g_{bc}=0$.  

Now we consider a smooth one-parameter family of curves $C_{\alpha}(\tau)$, 
parameterized by a proper time $\tau$ $(\tau_{1}\le\tau\le\tau_{2})$ such that 
for all $\alpha$ and $p,q\in M$, $C_{\alpha}(\tau_{1})=p$, 
$C_{\alpha}(\tau_{2})=q$, and $C_{0}$ is a geodesic or a classical path, along 
which a tangent vector field $v^{a}$ satisfies the geodesic equation
\beq
v^{a}\nabla_{a}v^{b}=\frac{{\rm d}^{2}x^{i}}{{\rm d}\tau^{2}}
+\Gamma^{i}_{jk}\frac{{\rm d}x^{j}}{{\rm d}\tau}
\frac{{\rm d}x^{k}}{{\rm d}\tau}=0
\label{geodesic}
\eeq  
where $\nabla_{a}v^{b}=\partial_{a}v^{b}+\Gamma^{b}_{ac}v^{c}$.  Let $\Sigma$ 
be a two-dimensional submanifold spanned by curves $C_{\alpha}(\tau)$ and 
we choose ($\tau$, $\alpha$) as coordinates of $\Sigma$.  The vector fields 
$v^{a}=(\partial/\partial\tau)^{a}$ and $w^{a}=(\partial/\partial\alpha)^{a}$ 
are then the tangent to the family of curves and the deviation vector 
representing the displacement to an infinitesimally nearby curve, 
respectively.  Here one notes that $w^{a}$ can be always chosen orthogonal to 
$v^{a}$ and vanishes at end-points to yield the boundary conditions, 
\beq 
w^{a}(\tau_{1})=w^{a}(\tau_{2})=0.
\label{bcwa}  
\eeq
Since $v^{a}$ and $w^{a}$ are coordinate vector fields, they commute to each 
other,
\beq
\pounds_{v}w^{a}=v^{b}\nabla_{b}w^{a}-w^{b}\nabla_{b}v^{a}={\rm 0}.
\label{comm}
\eeq    
 
In the stationary phase approximation where $|w^{a}|$ is infinitesimally 
small, one can expand the action (\ref{action}) around the geodesic 
$C_{0}$
\beq
S=S_{cl}+S^{(1)}[w(\tau)]+\frac{1}{2}S^{(2)}[w(\tau)]+\cdots,
\label{exp}
\eeq
where $S_{cl}=S|_{\alpha =0}$ is a classical action, and 
\footnote{For simplicity, we will 
parameterize the curve so that the Lagrangian $L$ is given as 
$L=(g_{ab}v^{a}v^{b})^{1/2}=1$ along the geodesic without loss of 
generality, since the action (\ref{action}) is parameterization independent.  
Here we need to put the condition $\alpha =0$ at the end of the calculations 
of $S^{(1)}[w(\tau)]$ and $S^{(2)}[w(\tau)]$.}
\bea
S^{(1)}[w(\tau)]
&=&\int_{\tau_{1}}^{\tau_{2}}{\rm d}\tau~w^{a}\nabla_{a}(v^{b}v_{b})^{1/2}
\nonumber\\
&=&\int_{\tau_{1}}^{\tau_{2}}{\rm d}\tau~v_{b}w^{a}\nabla_{a}v^{b}
\nonumber\\
&=&-\int_{\tau_{1}}^{\tau_{2}}{\rm d}\tau~w^{b}v^{a}\nabla_{a}v_{b}
\label{ds1}\\
S^{(2)}[w(\tau)]
&=&-\int_{\tau_{1}}^{\tau_{2}}{\rm d}\tau~w^{c}\nabla_{c}
(w_{b}v^{d}\nabla_{d}v^{b})
\nonumber\\
&=&-\int_{\tau_{1}}^{\tau_{2}}{\rm d}\tau~w^{c}w_{b}(\nabla_{c}
v^{d}\nabla_{d}v^{b}+v^{d}\nabla_{c}\nabla_{d}v^{b})
\nonumber\\
&=&-\int_{\tau_{1}}^{\tau_{2}}{\rm d}\tau~g_{ab}w^{a}(v^{c}\nabla_{c}
(v^{d}\nabla_{d}w^{b})+R_{cde}^{~~~b}v^{c}v^{e}w^{d})
\nonumber\\
&=&\int_{\tau_{1}}^{\tau_{2}}{\rm d}\tau~g_{ij}w^{i}\Lambda^{j}_{~k}w^{k}
\label{ds2}
\eea
where we have the Sturm-Liouville operator given as   
\beq
\Lambda^{i}_{~j}=-\delta^{i}_{~j}\frac{{\rm d}^{2}}{{\rm d}\tau^{2}}
-R_{kjl}^{~~~i}v^{k}v^{l},
\label{lambda}
\eeq
and we have used Eqs. (\ref{geodesic}) - (\ref{comm}) and 
the convention for the Riemann curvature tensor for any 
vector field $v^{a}$~\cite{wald84}
\beq
(\nabla_{a}\nabla_{b}-\nabla_{b}\nabla_{a})v^{c}=-R_{abd}^{~~~c}v^{d}.
\label{riemann}
\eeq

Now we consider a partition function~\cite{feynman65}
\beq
Z(q,\tau_{2};p,\tau_{1})=\int D[x^{i}(\tau)]e^{iS[x^{i}(\tau)]},
\label{feynman}
\eeq
which, in the stationary phase approximation, contains a widely 
oscillatory integral~\cite{witten89,dashen75,rajaraman82} and is thus 
given by contributions from the points of stationary phase.  Here one notes 
that the stationary points precisely construct the 
geodesic, along which the total energy is constant.  Since the above 
$S^{(1)}[w(\tau)]$ in Eq. (\ref{ds1}) vanishes due to the geodesic 
equation (\ref{geodesic}), by inserting Eq. (\ref{exp}) into 
Eq. (\ref{feynman}), one can obtain the partition function of the form,
\bea
Z(q,\tau_{2};p,\tau_{1})&=&e^{iS_{cl}}Z^{(2)}(q,\tau_{2};p,\tau_{1})
\label{zorig}\\
Z^{(2)}(q,\tau_{2};p,\tau_{1})&=&\int D[w(\tau)]e^{iS^{(2)}[w(\tau)]/2}.
\label{part2}
\eea   

Now we define a scalar product 
\beq
(w,w^{\prime})=\int_{\tau_{1}}^{\tau_{2}}{\rm d}\tau~g_{ij}w^{i}(\tau)
w^{\prime j}(\tau)
\label{product}
\eeq
to transform the path space of the integral (\ref{part2}) into a Hilbert space 
which is an external product of two spaces: three-dimensional space of the 
physical system and an infinite-dimensional Hilbert space of the continuous 
scalar functions on $[\tau_{1},\tau_{2}]$ vanishing at $\tau_{1}$ and 
$\tau_{2}$.  By choosing an orthonormal basis $\{u^{i}_{\alpha}(\tau)\}$ in 
this Hilbert space, one can have the deviation vector $w^{i}(\tau)$ in terms of 
superposition of $\{u^{i}_{\alpha}(\tau)\}$
\beq
w^{i}(\tau)=\sum_{\alpha =1}^{\infty}a^{\alpha}u^{i}_{\alpha} (\tau)
\label{witaus}
\eeq
so that, together with Eq. (\ref{product}), one can rewrite the second order
action (\ref{ds2}) as follows
\beq
S^{(2)}[w(\tau)]=\sum_{\alpha,\beta}c_{\alpha\beta}a^{\alpha}a^{\beta}
\label{caa}
\eeq
where
\beq
c_{\alpha\beta}=\int_{\tau_{1}}^{\tau_{2}}{\rm d}\tau~g_{ij}u^{i}_{\alpha}
\Lambda^{j}_{k}u^{k}_{\beta}=(u_{\alpha},\Lambda u_{\beta}).
\label{cab}
\eeq

Now in order to diagonalize the matrix $c_{\alpha\beta}$ we find an 
orthonormal basis of eigenfunctions $\{u^{i}_{\alpha}(\tau)\}$ of the operator 
$\Lambda$ with eigenvalues $\lambda_{\alpha}$ via the following eigenvalue 
equations, which is associated with the Morse 
theory~\cite{milnor63,wald84,morse34} and is also algebraically treated to 
compute the semiclassical partition function for the secondary 
Lagrangian~\cite{levit77},
\beq
-\Lambda^{i}_{~j}u^{j}_{\alpha}+\lambda_{\alpha}u^{i}_{\alpha}=
\frac{{\rm d}^{2}u^{i}_{\alpha}}{{\rm d}\tau^{2}}
+R_{kjl}^{~~~i}v^{k}v^{l}u^{j}_{\alpha}+\lambda_{\alpha}u^{i}_{\alpha}=0,
\label{morse}
\eeq
with 
\beq
u^{i}_{\alpha}(\tau_{1})=u^{i}_{\alpha}(\tau_{2})=0, ~~~~(i=1,2,3).
\label{bc2}
\eeq
With the above eigenvalues and eigenfunctions, one can obtain 
\beq
c_{\alpha\beta}=\lambda_{\alpha}(u_{\alpha},u_{\beta})=
\lambda_{\alpha}\delta_{\alpha\beta}
\label{cab2}
\eeq
to yield
\beq
S^{(2)}[w(\tau)]=\sum_{\alpha}\lambda_{\alpha}(a^{\alpha})^{2}
\label{s2lam}
\eeq
from which one can rewrite the second order partition function (\ref{part2}) 
as follows
\beq
Z^{(2)}(q,\tau_{2};p,\tau_{1})=J\prod_{\alpha=1}^{\infty}
\int_{-\infty}^{\infty}{\rm d}a^{\alpha}e^{i\lambda_{\alpha}
(a^{\alpha})^{2}/2}.
\label{z22}
\eeq
Here $J$ is the Jacobian defined as 
\beq
D[w(\tau)]=J\prod_{\alpha=1}^{\infty}{\rm d}a^{\alpha}
\label{dwt}
\eeq
and is independent of $w(\tau)$ due to the linearity of the 
transformation (\ref{witaus}), so that $J$ can be brought out of the integral 
symbol.

By taking the vanishing $\epsilon$ limit of the absolutely convergent 
integral, one can obtain 
\beq
\lim_{\epsilon\rightarrow 0}\int_{-\infty}^{\infty}{\rm d}a^{\alpha}
e^{i\lambda_{\alpha}(a^{\alpha})^{2}/2}
e^{-\epsilon (a^{\alpha})^{2}}=
e^{i\pi~{\rm sign}\lambda_{\alpha}/4}\left|\frac{2\pi}{\lambda_{\alpha}}
\right|^{1/2}
\eeq
to arrive at
\beq
Z^{(2)}(q,\tau_{2};p,\tau_{1})=Je^{i\pi\sum_{\alpha}{\rm sign}
\lambda_{\alpha}/4}\prod_{\alpha}^{\infty}
\left|\frac{2\pi}{\lambda_{\alpha}}\right|^{1/2}.
\label{z2j}
\eeq
Here note that the phase factor is proportional to 
$\sum_{\alpha}{\rm sign}\lambda_{\alpha}$, which is associated with the eta 
invariant of Atiyah~\cite{atiyah75,witten89}
\beq
\eta=\frac{1}{2}\lim_{s\rightarrow 0}~{\rm sign}\lambda_{\alpha}
|\lambda_{\alpha}|^{-s}.
\label{eta}
\eeq

Now we relate our geodesic to geodesic of a free particle with the same 
constant energy $E$ to obtain a closed form of the absolute value in Eq. 
(\ref{z2j}).  Recalling that the Jacobian $J$ remains unchanged for the 
unitary transformation 
$\{ u_{\alpha}(\tau)\}\rightarrow \{\hat{u}_{\alpha}(\tau)\}$~\cite{levit77}, 
for the free particle with the metric $\hat{g}_{ij}=\frac{1}{2}mE\delta_{ij}$ 
which can be obtained from Eq. (\ref{metric}), one can have the partition 
function
\beq
\hat{Z}^{(2)}(q,\tau_{2};p,\tau_{1})=Je^{i\pi\sum_{\alpha}~{\rm sign}
\hat{\lambda}_{\alpha}/4}\prod_{\alpha}^{\infty}
\left|\frac{2\pi}{\hat{\lambda}_{\alpha}}\right|^{1/2},
\label{z2j0}
\eeq  
whose corresponding eigenvalue equations can be described as 
\beq
-\hat{\Lambda}^{i}_{~j}\hat{u}^{j}_{\alpha}+\hat{\lambda}_{\alpha}
\hat{u}^{i}_{\alpha}=
\frac{{\rm d}^{2}\hat{u}^{i}_{\alpha}}{{\rm d}\tau^{2}}
+\hat{\lambda}_{\alpha}\hat{u}^{i}_{\alpha}=0,
\label{morsefree}
\eeq
with 
\beq
\hat{u}^{i}_{\alpha}(\tau_{1})=\hat{u}^{i}_{\alpha}(\tau_{2})=0, ~~~~(i=1,2,3).
\label{bcfree}
\eeq
Combination of Eqs. (\ref{z2j}) and (\ref{z2j0}) yields
\beq
Z^{(2)}(q,\tau_{2};p,\tau_{1})=e^{i\pi\sum_{\alpha}~({\rm sign}
~\lambda_{\alpha}-{\rm sign}~\hat{\lambda}_{\alpha})/4}
\left|\frac{\prod_{\alpha}\hat{\lambda}_{\alpha}}
{\prod_{\alpha}\lambda_{\alpha}}\right|^{1/2}
\hat{Z}^{(2)}(q,\tau_{2};p,\tau_{1}).
\label{z2jz2jfree}
\eeq

Now we consider a Jacobi equation to express the absolute value of the ratio 
in Eq. (\ref{z2jz2jfree}) in terms of the initial data at the starting 
point $p=x^{i}(\tau_{1})$, by introducing a smooth one-parameter family 
of geodesics $\gamma_{\alpha}(\tau)$ on the manifold $M$.  Here one can vary 
the parameter $\alpha\in R$ by infinitesimally changing the direction of the 
initial velocity $v^{i}(\tau_{1})={\rm d}x^{i}/{\rm d}\tau (\tau_{1})$ at 
$p$, and also one can choose $\alpha$ and $\tau$ as coordinates of 
a submanifold $\Sigma_{\gamma}$ spanned by the geodesics 
$\gamma_{\alpha}(\tau)$ on $M$.  Along the geodesic $\gamma_{0}$, one can 
have tangent vector fields $v^{a}$ and deviation vector field $w^{a}$ which 
points to an 
infinitesimally nearby geodesic and vanishes at $p$, namely, 
$w^{a}(\tau_{1})=0$, so that one can have the relative acceleration of the 
displacement to an infinitesimally nearby geodesic,\footnote{Differently from 
Eqs. (\ref{ds1}) and (\ref{ds2}), one can exploit the fact that all curves 
involved in $a^{a}$ are geodesics, so that one can use the identity 
$w^{c}\nabla_{c}(v^{b}\nabla_{b}v^{a})=0$.} 
\beq
a^{a}=v^{c}\nabla_{c}(v^{b}\nabla_{b}w^{a})
=-R_{bcd}^{~~~a}v^{b}v^{d}w^{c},
\label{accel}
\eeq
to yield the geodesic deviation equation
\beq
v^{c}\nabla_{c}(v^{b}\nabla_{b}w^{a})
+R_{bcd}^{~~~a}v^{b}v^{d}w^{c}
=\frac{{\rm d}^{2}w^{i}_{\alpha}}{{\rm d}\tau^{2}}
+R_{klm}^{~~~i}v^{k}v^{m}w^{l}=-\Lambda^{i}_{~j}w^{j}=0,
\label{jacobieqn}
\eeq
which is also known as a Jacobi equation and its solution $w^{a}$ is named 
a Jacobi field on the geodesic $\gamma_{0}$ whose tangent is $v^{a}$.  Since the 
above Jacobi equation (\ref{jacobieqn}) is a linear differential 
equation, the Jacobi field $w^{i}(\tau)$ depends linearly on the inertial 
data $w^{i}(\tau_{1})=p$ and ${\rm d}w^{i}/{\rm d}\tau (\tau_{1})$ at the 
starting point $p$ to yield
\beq
w^{i}(\tau)=T^{i}_{~j}(\tau)\frac{{\rm d}w^{j}}{{\rm d}\tau}(\tau_{1}),
\label{tijtau}
\eeq
where 
\beq
T^{i}_{~j}(\tau_{1})=0,~~~
\frac{{\rm d}T^{i}_{~j}}{{\rm d}\tau}(\tau_{1})=\delta^{i}_{~j}
\label{tijinis}
\eeq
and $T^{i}_{~j}(\tau)$ can be defined as\footnote{
Since the coordinates $x^{i}(\tau)$ and the velocity 
$v^{i}(\tau)={\rm d}x^{i}/{\rm d}\tau (\tau)$ are independent variables at 
the same time, say $\tau_{1}$, one can easily check that the definition 
(\ref{tijdef}) satisfies Eq. (\ref{tijinis}).} 
\beq
T^{i}_{~j}(\tau)=\frac{{\rm d}x^{i}(\tau)}{{\rm d}v^{j}(\tau_{1})}.
\label{tijdef}
\eeq
Substituting Eq. (\ref{tijtau}) into Eq. (\ref{jacobieqn}), one can rewrite 
the Jacobi equation in terms of $T^{i}_{~j}$ as
\beq
\frac{{\rm d}^{2}T^{i}_{~j}}{{\rm d}\tau^{2}}
+R_{klm}^{~~~i}v^{k}v^{m}T^{l}_{~j}=-\Lambda^{i}_{~l}T^{l}_{~j}=0.
\label{jacobieqnt}
\eeq 
Similarly, for the free particle one can obtain
\bea
\frac{{\rm d}^{2}\hat{T}^{i}_{~j}}{{\rm d}\tau^{2}}
&=&-\hat{\Lambda}^{i}_{~l}\hat{T}^{l}_{~j}=0,
\label{jacobieqntfree}\\
\hat{T}^{i}_{~j}(\tau_{1})&=&0,~~~
\frac{{\rm d}\hat{T}^{i}_{~j}}{{\rm d}\tau}(\tau_{1})=\delta^{i}_{~j}
\label{tijinisfree}
\eea 
from which the ratio in Eq. (\ref{z2jz2jfree}) can be rewritten 
as~\cite{levit7799}
\beq
\left|\frac{\prod_{\alpha}\hat{\lambda}_{\alpha}}
{\prod_{\alpha}\lambda_{\alpha}}\right|=
\left|\frac{{\rm det}~\hat{T}^{i}_{~j}(\tau_{2})}{{\rm det}~T^{i}_{~j}
(\tau_{2})}\right|.
\label{ratiott}
\eeq
On the other hand, the partition function for a free particle is given 
as~\cite{papa75}
\beq
\hat{Z}^{(2)}(q,\tau_{2};p,\tau_{1})=(2\pi i)^{-3/2}
\left(\frac{{\rm det}~\hat{g}_{ij}(\tau_{1})}{{\rm det}~\hat{T}^{i}_{~j}
(\tau_{2})}\right)^{1/2}.
\label{z2part}
\eeq
Using the above results (\ref{ratiott}) and (\ref{z2part}), the second order 
partition function (\ref{z2jz2jfree}) can be rewritten as
\beq
Z^{(2)}(q,\tau_{2};p,\tau_{1})=(2\pi i)^{-3/2}
e^{i\pi\sum_{\alpha}({\rm sign}\lambda_{\alpha}
-{\rm sign}\hat{\lambda}_{\alpha})/4}
\left|\frac{{\rm det}~\hat{g}_{ij}(\tau_{1})}{{\rm det}~T^{i}_{~j}(\tau_{2})}
\right|^{1/2}
\label{z2dethat}
\eeq
to yield 
\beq
Z^{(2)}(q,\tau_{2};p,\tau_{1})=(2\pi i)^{-3/2}
e^{i\pi\sum_{\alpha}({\rm sign}\lambda_{\alpha}
-{\rm sign}\hat{\lambda}_{\alpha})/4}
\left|{\rm det}~\hat{g}_{ij}(\tau_{1})\frac{\partial v^{j}(\tau_{1})}
{\partial x^{k}(\tau_{2})}\right|^{1/2}.
\label{z2detvx}
\eeq

Now we consider the Hamilton-Jacobi theory~\cite{goldstein80} where classical 
conjugate momentum $p_{i}^{cl}(\tau)$ corresponding to $x^{i}(\tau)$ 
in the action (\ref{action}) is given as 
\beq
p_{i}^{cl}(\tau)=\frac{\partial L_{cl}}{\partial v^{i}}=g_{ij}v^{j},~~~
p_{i}^{cl}(\tau_{1})=\frac{\partial S_{cl}}{\partial x^{i}(\tau_{1})}
\label{cojmon}
\eeq
from which we can obtain
\bea
\hat{g}_{ij}(\tau_{1})&=&\frac{\partial p_{i}^{cl}(\tau_{1})}
{\partial v^{j}(\tau_{1})},
\nonumber\\
\frac{\partial v^{i}(\tau_{1})}{\partial x^{j}(\tau_{2})}
&=&\frac{1}{\hat{g}_{ik}(\tau_{1})}
\frac{\partial p_{k}^{cl}(\tau_{1})}{\partial x^{j}(\tau_{2})}
=\frac{1}{\hat{g}_{ik}(\tau_{1})}
\frac{\partial^{2}S_{cl}}{\partial x^{k}(\tau_{1})\partial x^{j}(\tau_{2})}
\label{gvx}
\eea
to, together with Eqs. (\ref{zorig}) and (\ref{z2detvx}), yield the desired 
semiclassical partition function
\beq
Z(q,\tau_{2};p,\tau_{1})=(2\pi i)^{-3/2}
e^{i\pi\sum_{\alpha}({\rm sign}\lambda_{\alpha}
-{\rm sign}\hat{\lambda}_{\alpha})/4}
e^{iS_{cl}}\left|{\rm det}\frac{\partial^{2}S_{cl}}{\partial x^{i}(\tau_{1})
\partial x^{k}(\tau_{2})}\right|^{1/2},
\label{final}
\eeq
where the determinant involved here is known as the Van Vleck 
determinant~\cite{vleck28}.  Here ${\rm sign}\hat{\lambda}_{\alpha}$ is easily 
checked to be positive and ${\rm sign}\lambda_{\alpha}$ could be positive 
or negative, so that the corresponding additional phase factor in the 
semiclassical partition function (\ref{final}) is given by 
$e^{-i{\cal N}\pi/2}$ where ${\cal N}$ is the number of negative eigenvalues, 
counted with their multiplicity~\cite{morse34,levit77}, of the operator 
$\Lambda$ in the eigenvalue equation (\ref{morse}) of the Morse theory.  

In conclusion, we have explicitly derived the semiclassical partition 
function in the framework of the Morse theory, where we can introduce the 
Jacobi fields and the Hamilton-Jacobi theory to yield the phase factor of the 
semiclassical partition function associated with the eta invariant of Atiyah.  
It will be interesting, through further investigation, to study conjugate 
points~\cite{milnor63} in the framework of this semiclassical approach 
to differential geometry.

\vskip 0.5cm
The author would like to thank the hospitality of the C.N. Yang 
Institute for Theoretical Physics, SUNY at Stony Brook and the CIT-USC 
Center for Theoretical Physics, University of Southern California, where a 
part of this work has been done.  He also would like to thank Itzhak Bars, 
John Milnor, Chiara Nappi, Peter van Nieuwenhuizen and Edward Witten for 
helpful discussions and concerns.  This work is supported in part by Grant No. 
2000-2-11100-002-5 from the Basic Research Program of the Korea Science and 
Engineering Foundation.


\end{document}